\documentstyle[epsfig]{mn2e}

\begin{document}

\title{The angular power spectrum of NVSS radio galaxies}

\author[Chris Blake, Pedro G. Ferreira and Julian Borrill]{Chris
Blake$^{\,1,2,}$\footnotemark, Pedro G. Ferreira$^{\,2}$ and Julian
Borrill$^{\,3}$ \\ \\ $^1$ School of Physics, University of New South
Wales, Sydney, NSW 2052, Australia \\ $^2$ Astrophysics, University of
Oxford, Keble Road, Oxford, OX1 3RH, UK \\ $^3$ National Energy
Research Scientific Computing Centre, LBNL, Berkeley, CA, USA}

\maketitle

\begin{abstract}
We measure the angular power spectrum $C_\ell$ of radio galaxies in
the NRAO VLA Sky Survey (NVSS) using two independent methods: direct
spherical harmonic analysis and maximum likelihood estimation.  The
results are consistent and can be understood using models for the
spatial matter power spectrum and for the redshift distribution of
radio galaxies at mJy flux-density levels.  A good fit to the angular
power spectrum can only be achieved if radio galaxies possess high
bias with respect to mass fluctuations; by marginalizing over the
other parameters of the model we derive a $68\%$ confidence interval
$1.53 < b_0 \times \sigma_8 < 1.87$, where $b_0$ is the linear bias
factor for radio galaxies and $\sigma_8$ describes the normalization
of the matter power spectrum.  Our models indicate that the majority
of the signal in the NVSS $C_\ell$ spectrum is generated at low
redshifts $z \la 0.1$.  Individual redshifts for the NVSS sources are
thus required to alleviate projection effects and probe directly the
matter power spectrum on large scales.
\end{abstract}

\section{Introduction}
\renewcommand{\thefootnote}{\fnsymbol{footnote}}
\setcounter{footnote}{1}
\footnotetext{E-mail: chrisb@phys.unsw.edu.au}

\begin{figure*}
\center
\epsfig{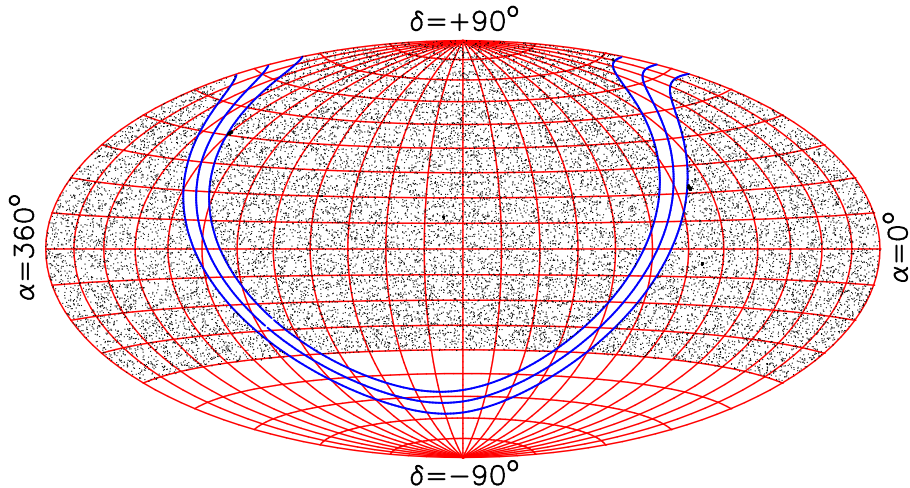}
\caption{NVSS sources with $S_{\rm 1.4 \, GHz} > 200$ mJy in an
equal-area projection.  The Galactic plane and Galactic latitudes $\pm
5^\circ$ are also plotted; sources within this region are masked from
our large-scale structure analysis as many are Galactic in origin.}
\label{fignvss}
\end{figure*}

Active Galactic Nuclei (AGN) mapped in radio waves are an interesting
probe of large-scale structure.  They can be routinely detected out to
very large redshift ($z \sim 4$) over wide areas of the sky and hence
delineate the largest structures and their evolution over cosmic
epoch.  Radio emission is insensitive to dust obscuration and radio
AGN are effective tracers of mass: they are uniformly hosted by
massive elliptical galaxies and have been shown to trace both clusters
(Hill \& Lilly 1991) and superclusters (Brand et al. 2003).

The current generation of wide-area radio surveys such as Faint Images
of the Radio Sky at Twenty centimetres (FIRST; Becker, White \&
Helfand 1995) and the NRAO VLA Sky Survey (NVSS; Condon et al. 1998)
contain radio galaxies in very large numbers ($\sim 10^6$) and have
allowed accurate measurements of the imprint of radio galaxy angular
clustering.  These patterns are considerably harder to detect in radio
waves than in optical light due to the huge redshift range that is
probed.  Whilst this provides access to clustering on the largest
scales, it also washes out much of the angular clustering signal
through the superposition of unrelated redshift slices.  The angular
correlation function was measured for FIRST by Cress et al. (1996) and
Magliocchetti et al. (1998) and for NVSS by Blake \& Wall (2002a) and
Overzier et al. (2003).

The NVSS radio survey, covering $\sim 80$ per cent of the sky, permits
the measurement of fluctuations over very large angles.  Blake \& Wall
(2002b) detected the imprint of the cosmological velocity dipole in
the NVSS surface density, in a direction consistent with the Cosmic
Microwave Background (CMB) dipole.  In this study we measure the
angular power spectrum, $C_\ell$, of the radio galaxy distribution
(Baleisis et al. 1998).  This statistic represents the source
surface-density field as a sum of sinusoidal angular density
fluctuations of different wavelengths, using the spherical harmonic
functions.  The angular power spectrum is sensitive to large-angle
fluctuations and hence complements the measurement of the angular
correlation function, $w(\theta)$, at small angles.

Measurement of the $C_\ell$ spectrum has some advantages in comparison
with $w(\theta)$.  Firstly, the error matrix describing correlations
between multipoles $\ell$ has a very simple structure, becoming
diagonal for a complete sky.  This is not the case for the separation
bins in a measurement of $w(\theta)$: even for a full sky, an
individual galaxy appears in many separation bins, automatically
inducing correlations between those bins.  Secondly, there is a
natural relation between the angular power spectrum and the spatial
power spectrum of density fluctuations, $P(k)$.  This latter quantity
provides a very convenient means of describing structure in the
Universe for a number of reasons.  Firstly its primordial form is
produced by models of inflation, which prescribe the initial pattern
of density fluctuations $\delta\rho/\rho$.  Furthermore, in linear
theory for the growth of perturbations, fluctuations described by
different wavenumbers $k$ evolve independently, enabling the model
power spectrum to be easily scaled with redshift.  The physics of
linear perturbations are hence more naturally described in Fourier
space.

In contrast, the angular correlation function is more easily related
to the spatial correlation function $\xi(r)$, the Fourier transform of
$P(k)$.  Correlation functions more naturally serve to describe the
real-space profile of collapsing structures evolving out of the linear
regime.  We emphasize that although the two functions $C_\ell$ and
$w(\theta)$ are {\it theoretically} equivalent -- linked by a Legendre
transform -- this is not true in an {\it observational} sense.  For
example, $w(\theta)$ can only be successfully measured for angles up
to a few degrees, but $C_\ell$ depends on $w(\theta)$ at {\it all}
angles.

We derive the angular power spectrum using two independent methods.
Firstly we apply a direct spherical harmonic estimator following
Peebles (1973).  Secondly, we use maximum likelihood estimation,
commonly employed for deriving the angular power spectra of the CMB
temperature and polarization maps.  These two methods are described in
Section \ref{secmeth}.  We find that these two approaches yield very
similar results (Section \ref{secres}), which is unsurprising given
the wide sky coverage of the NVSS.  In Section \ref{secpk} we
interpret the NVSS angular power spectrum in terms of the underlying
spatial power spectrum of mass fluctuations and the radial
distribution of radio sources.  Finally in Section \ref{secbias} we
employ these models to derive the linear bias factor of NVSS radio
galaxies by marginalizing over the other model parameters.

\section{The NVSS radio survey}
\label{secobs}

The 1.4 GHz NRAO VLA Sky Survey (NVSS; Condon et al. 1998) was
performed at the Very Large Array over the period 1993 to 1996 and
covers the sky north of declination $-40^\circ$.  The source catalogue
contains $\approx 1.8 \times 10^6$ entries and is 99 per cent complete
at integrated flux density $S_{\rm 1.4 \, GHz} = 3.5$ mJy.  The full
width at half-maximum of the synthesized beam is 45 arcsec; the
majority of radio sources are thus unresolved.  The relatively broad
NVSS beam yields excellent surface brightness sensitivity and
photometric completeness.

Before analyzing the survey for large-scale structure we imposed
various angular masks.  Firstly we excluded catalogue entries within
$5^\circ$ of the Galactic plane, many of which are Galactic in origin
(mostly supernova remnants and HII regions).  The contribution of
foreground Galactic sources at latitudes $|b| > 5^\circ$ is
negligible.  We also placed 22 masks around bright local extended
radio galaxies contributing large numbers of catalogue entries, as
described in Blake \& Wall (2002a).  Figure \ref{fignvss} plots the
remaining sources with $S_{\rm 1.4 \, GHz} > 200$ mJy.  The remaining
NVSS geometry corresponds to 75 per cent of the celestial sphere.

Blake \& Wall (2002a) demonstrated that the NVSS suffers from
systematic gradients in surface density at flux-density levels at
which it is complete.  These gradients -- corresponding to a $\sim 5$
per cent variation in surface density at a threshold of 3 mJy -- are
entirely unimportant for the vast majority of applications of this
catalogue.  However, they have a significant influence on the faint
imprint of large-scale structure.  If left uncorrected a distortion of
the measured angular power spectrum would result, because the harmonic
coefficients would need to reproduce the systematic gradients as well
as the fluctuations due to clustering.

Blake \& Wall (2002a) found that these surface gradients are only
significant at fluxes $S_{\rm 1.4 \, GHz} < 10$ mJy.  At brighter
fluxes the survey is uniform to better than 1 per cent -- sufficient
to allow the detection of the anticipated cosmological velocity dipole
(Blake \& Wall 2002b).  Hence we simply restricted our $C_\ell$
analysis to fluxes $S_{\rm 1.4 \, GHz} > 10$ mJy.  The NVSS source
surface density at this threshold is $\sigma_0 = 16.9$ deg$^{-2}$.

We note that the broadness of the radio luminosity function ensures
that the projected clustering properties of radio galaxies are not a
strong function of flux density in the range 3 mJy $< S_{\rm 1.4 \,
GHz} <$ 50 mJy, as verified by the correlation function analyses of
Blake \& Wall (2002a) and Overzier et al. (2003).  In this
flux-density range, the redshift distribution of the radio galaxies
does not change significantly.

\section{Estimating the angular power spectrum: methods}
\label{secmeth}

\subsection{Definition of the angular power spectrum}

A distribution of galaxies on the sky can be generated in two
statistical steps.  Firstly, a density field $\sigma(\theta,\phi)$ is
created; this may be described in terms of its spherical harmonic
coefficients $a_{\ell m}$:
\begin{equation}
\sigma(\theta,\phi) = \sum_{\ell=0}^{\infty} \sum_{m=-\ell}^{+\ell} \,
a_{\ell m} \, Y_{\ell m}(\theta,\phi)
\label{eqsigylm}
\end{equation}
\noindent where $Y_{\ell m}$ are the usual spherical harmonic
functions.  Secondly, galaxy positions are generated in a Poisson
process as a (possibly biased) realization of this density field.

The angular power spectrum $C_\ell$ prescribes the spherical harmonic
coefficients in the first step of this model.  It is defined over many
realizations of the density field by
\begin{equation}
<|a_{\ell m}|^2> = C_\ell
\label{eqcldef}
\end{equation}
The assumption of isotropy ensures that $<|a_{\ell m}|^2>$ is a
function of only $\ell$, not $m$.

The angular power spectrum $C_\ell$ is theoretically equivalent to the
angular correlation function $w(\theta)$ as a description of the
galaxy distribution.  The two quantities are connected by the
well-known relation
\begin{equation}
C_\ell = 2 \pi \, \sigma_0^2 \int_{-1}^{+1} \, w(\theta) \,
P_\ell(\cos{\theta}) \, d(\cos{\theta})
\label{eqwtocl}
\end{equation}
where $\sigma_0$ is the source surface density and $P_\ell$ is the
Legendre polynomial.  However, the angular scales on which the
signal-to-noise is highest are very different for each statistic.
$w(\theta)$ can only be measured accurately at small angles up to a
few degrees (Blake \& Wall 2002a), beyond which Poisson noise
dominates.  By contrast, $C_\ell$ for galaxies has highest
signal-to-noise at small $\ell$, corresponding to large angular scales
$\theta \sim 180^\circ/\ell$.  Hence the two statistics are
complementary, the $C_\ell$ spectrum probing fluctuations on the
largest angular scales.

\subsection{Spherical harmonic estimation of $C_\ell$}
\label{secestharm}

Peebles (1973) presented the formalism of spherical harmonic analysis
of a galaxy distribution over an incomplete sky (for refinements see
e.g. Wright et al. 1994; Wandelt, Hivon \& Gorski 2000).  For a
partial sky, a spherical harmonic analysis is hindered by the fact
that the spherical harmonics are not an orthonormal basis, which
causes the measured coefficients $a_{\ell m}$ to be statistically
correlated, entangling different multipoles of the underlying $C_\ell$
spectrum.  However, for the case of a survey covering $\sim 80\%$ of
the sky, the repercussions (discussed below) are fairly negligible,
implying shifts and correlations in the derived power spectrum that
are far smaller than the error bars.  We employed the original method
of Peebles (1973) with only one small correction for sample variance.
In Section \ref{secmaxlik} we compare spherical harmonic analysis with
the technique of maximum likelihood estimation and find that the two
methods yield results in good agreement.

The spherical harmonic coefficients of the density field may be
estimated by summing over the $N$ galaxy positions
$(\theta_i,\phi_i)$:
\begin{equation}
A_{\ell m} = \sum_{i=1}^{N} \, Y_{\ell m}^*(\theta_i,\phi_i)
\label{eqalm}
\end{equation}
For an incomplete sky, these values need to be corrected for the
unsurveyed regions, so that an estimate of $C_\ell$ is
\begin{equation}
C_{\ell m}^{\rm obs} = \frac{|A_{\ell m} - \sigma_0 \, I_{\ell
m}|^2}{J_{\ell m}} - \sigma_0
\label{eqclpeeb}
\end{equation}
(Peebles 1973 equation 50) where $\sigma_0 = N/\Delta\Omega$ and
\begin{equation}
I_{\ell m} = \int_{\Delta \Omega} \, Y_{\ell m}^* \, d\Omega
\end{equation}
\begin{equation}
J_{\ell m} = \int_{\Delta \Omega} \, |Y_{\ell m}|^2 \, d\Omega
\label{eqjlm}
\end{equation}
where the integrals are over the survey area $\Delta \Omega$, and are
determined in our analysis by numerical integration.  The final term
in equation \ref{eqclpeeb} corrects for the finite number of discrete
sources: for a full sky ($I_{\ell m} = 0, J_{\ell m} = 1$) we expect
$<|A_{\ell m}|^2> = \sigma_0$ in the absence of clustering, i.e. this
is the power spectrum of the shot noise.  Note that $\sigma_0$ is the
apparent source density $N/\Delta \Omega$, not the average over an
imagined ensemble of catalogues.

We determined the angular power spectrum for the $\ell$th multipole,
$C_\ell^{\rm obs}$, by averaging equation \ref{eqclpeeb} over $m$.
Because the density field is real rather than complex, $C_{\ell
m}^{\rm obs} = C_{\ell,-m}^{\rm obs}$, resulting in $\ell+1$
independent measurements of $C_\ell$:
\begin{equation}
C_\ell^{\rm obs} = \frac{\sum_{m=0}^\ell \, C_{\ell m}^{\rm
obs}}{\ell+1}
\label{eqclobs}
\end{equation}
There is no need for us to use the modified weighting formula of
Peebles (1973 equation 53).  In our case, $J_{\ell m}$ does not vary
significantly with $m$.  We verified that the modified weighting
formula produced indistinguishable results.

One consequence of the partial sky is to ``mix'' the harmonic
coefficients such that the measured angular power spectrum at $\ell$
depends on a range of $C_{\ell'}$ around $\ell' = \ell$:
\begin{equation}
<C_\ell^{\rm obs}> = \sum_{\ell'} \, C_{\ell'} \, R_{\ell\ell'}
\end{equation}
The angled brackets refer to an imagined averaging over many
realizations of density fields generated by $C_\ell$, in accordance
with equation \ref{eqcldef}.  Peebles showed that $\sum_{\ell'}
R_{\ell\ell'} = 1$; i.e. mixing does not spuriously enhance the
measured power (this is accomplished by the factor $J_{\ell m}$ in
equation \ref{eqclpeeb}).

\begin{figure}
\center
\epsfig{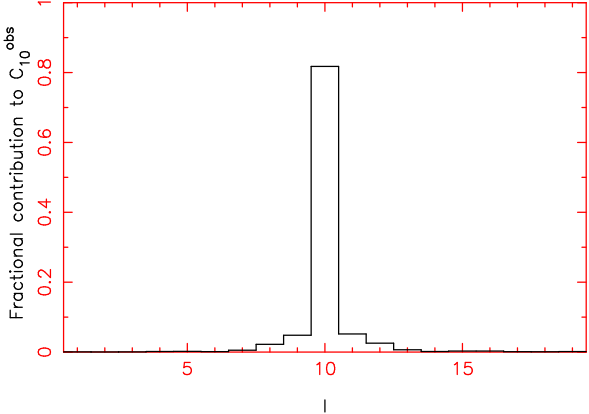}
\caption{The range of $\ell$ probed by a measurement of $C_\ell^{\rm
obs}$ at $\ell = 10$.  The incomplete sky means that $C_{10}^{\rm
obs}$ is not just sensitive to $C_{10}$, but also to the surrounding
$C_\ell$'s.  However, the large fraction of sky covered means that
this mixing is small.}
\label{figrll}
\end{figure}

For a complete sky, $R_{\ell\ell'} = \delta_{\ell\ell'}$, where
$\delta_{mn} = 1$ ($m = n$) or 0 ($m \ne n$).  For a partial sky, the
matrix $R_{\ell\ell'}$ can be computed from the geometry of the
surveyed region (Hauser \& Peebles 1973).  Figure \ref{figrll}
illustrates the result for the NVSS for $\ell = 10$ (computed using
Hauser \& Peebles 1973 equation 12).  The NVSS covers a sufficiently
large fraction of the sky (75 per cent) that mixing only occurs at the
$\sim 15$ per cent level and can be neglected because the underlying
$C_\ell$ spectrum is smooth:
\[ <C_\ell^{\rm obs}> \approx C_\ell \]
We checked the NVSS $R_{\ell\ell'}$ matrix for other multipoles $\ell$
and found very similar results.  This argument ensures that the
measured multipoles are statistically independent to a good
approximation.

The statistical error on the estimator of equation \ref{eqclpeeb} is
\begin{equation}
\sigma(C_{\ell m}^{\rm obs}) \approx (\sigma_0 + C_\ell) \sqrt{1 +
\delta_{m0}}
\label{eqclsig}
\end{equation}
(Peebles 1973 equation 81).  There are two components of the error:
\begin{itemize}
\item ``Shot noise'' ($\sigma_0$) because the number of discrete
objects is finite and therefore does not perfectly describe the
underlying density field.
\item ``Cosmic variance'' ($C_\ell$) because even with perfect
sampling of the density field, there are only a finite number of
harmonics associated with the $\ell$th multipole.
\end{itemize}

The error for the $m = 0$ case in equation \ref{eqclsig} is greater
because $A_{\ell 0}$ is purely real, rather than complex.  In the
latter case, we are averaging over the real and imaginary parts of
$A_{\ell m}$, two independent estimates of $C_\ell$, which reduces the
overall statistical error by a factor $\sqrt{2}$.  For a partial sky
equation \ref{eqclsig} is an approximation, because the variance of
multipoles of given $\ell$ depends on the underlying power spectrum at
$\ell' \ne \ell$.  As discussed above, this effect is negligible for
the NVSS.

The averaging over $m$ (equation \ref{eqclobs}) decreases the error in
the observation.  Combining the errors of equation \ref{eqclsig},
assuming estimates at different $m$ are statistically independent, the
resulting error in $C_\ell^{\rm obs}$ is
\begin{equation}
\sigma(C_\ell^{\rm obs}) \approx (\sigma_0 + C_\ell)
\frac{\sqrt{\ell+2}}{\ell+1}
\label{eqclerr}
\end{equation}
We used Monte Carlo simulations to verify that equation \ref{eqclerr}
produced results within 5 per cent of the true error for all relevant
multipoles.

Our only addition to the formalism of Peebles (1973) was to increase
the total variance on the estimate of $C_\ell$ by a factor $1/f_{\rm
sky}$, where $f_{\rm sky} = \Delta \Omega/4\pi$ is the fraction of sky
covered (i.e. multiply equation \ref{eqclerr} by $1/\sqrt{f_{\rm
sky}}$).  This correction factor was motivated by Scott, Srednicki and
White (1994) as a fundamental property of sample variance for a
partial sky, and is part of the standard CMB formalism (e.g. Bond,
Efstathiou \& Tegmark 1997).  For the NVSS geometry, $f_{\rm sky} =
0.75$, thus this correction corresponds to a $\sim 10$ per cent
increase in the error.

\subsection{Correction for multiple-component sources}
\label{secmult}

Radio sources have complex morphologies and large linear sizes (up to
and exceeding 1 Mpc).  A radio-source catalogue such as the NVSS will
contain entries which are different components of the same galaxy (for
example, the two radio lobes of a ``classical double'' radio galaxy).
The broad angular resolution of the NVSS beam leaves over 90 per cent
of radio sources unresolved; however, the remaining multiple-component
sources have a small but measurable effect on the angular power
spectrum.

It is relatively simple to model the effect of multiple-component
sources on the estimator for $C_\ell$ described in Section
\ref{secestharm}.  The relevant angular scales ($\ell < 100$) are much
bigger than any component separation, and equation \ref{eqalm} can be
replaced by
\[ A_{\ell m} = \sum_{i=1}^{N_{\rm gal}} c_i Y_{\ell m}^*(i) \]
where $N_{\rm gal}$ is the total number of galaxies and $c_i$ is the
number of components of the $i$th galaxy.  Thus the quantity
\[ <A_{\ell m}> = N_{\rm gal} \, \overline{c} <Y_{\ell m}> = N <Y_{\ell m}> \]
is unchanged by the presence of multiple components ($\overline{c}$
denotes the average number of components per galaxy, and $N =
\overline{c} \times N_{\rm gal}$ is the total number of catalogue
entries, as in equation \ref{eqalm}).  But $C_{\ell m}^{\rm obs}$ in
equation \ref{eqclpeeb} depends on
\begin{eqnarray}
<|A_{\ell m}|^2> \hspace{-3mm} &=& \hspace{-3mm} N_{\rm gal} \,
\overline{c^2} <|Y_{\ell m}|^2> + N_{\rm gal}^2 \, \overline{c}^2
<Y_{\ell m}> <Y_{\ell m}^*> \nonumber \\ &=& \hspace{-3mm}
\frac{\overline{c^2}}{\overline{c}} \, N <|Y_{\ell m}|^2> \, + \, N^2
<Y_{\ell m}> <Y_{\ell m}^*> \nonumber
\end{eqnarray}
Multiple-component sources only affect the first term in this
expression, producing an offset in the $C_\ell$ spectrum
\[ \Delta C_\ell = \frac{\Delta (<|A_{\ell m}|^2>)}{J_{\ell m}} = \left( \frac{\overline{c^2}}{\overline{c}} - 1 \right) \frac{N <|Y_{\ell m}|^2>}{J_{\ell m}} \]
But $J_{\ell m} = <|Y_{\ell m}|^2> \Delta \Omega$ from equation
\ref{eqjlm}, and this expression simplifies to an offset independent
of $\ell$:
\[ \Delta C_\ell = \left( \frac{\overline{c^2}}{\overline{c}} - 1 \right) \sigma_0 \]
Most multiple-component sources in the NVSS catalogue are double radio
sources.  Let a fraction $e \ll 1$ of the radio galaxies be doubles.
Then $\overline{c} = 1 + e$ and $\overline{c^2} = 1 + 3e$, thus the
constant offset may be written
\begin{equation}
\Delta C_\ell \approx 2 \, e \, \sigma_0
\label{eqcldoub}
\end{equation}
We can deduce $e = 0.07 \pm 0.005$ from the form of the NVSS angular
correlation function $w(\theta)$ at small angles $\theta < 0.1^\circ$,
where double sources dominate the close pairs (see Blake \& Wall 2002a
and also Section \ref{secalm}).  This correction was applied to the
measured NVSS $C_\ell$ spectrum and successfully removed the small
systematic offset in $C_\ell$ at high $\ell$.

\subsection{Maximum likelihood estimation of $C_\ell$}
\label{secmaxlik}

A sophisticated suite of analytical tools has been developed by the
CMB community for deriving the angular power spectra of the observed
CMB temperature and polarization maps.  These methods can also be
exploited to analyze galaxy data (see for example Efstathiou \& Moody
2001, Huterer, Knox \& Nichol 2001 and Tegmark et al. 2002).  In this
approach the power spectrum is determined using an iterative maximum
likelihood analysis, in contrast to the direct estimator discussed in
Section \ref{secestharm}.  The likelihood is a fundamental statistical
quantity, and this analysis method permits straightforward control of
such issues as edge effects, noise correlations and systematic errors.

The starting point for maximum likelihood estimation (MLE) is Bayes'
theorem
\[ P(\alpha|DI) \propto P(\alpha|I) \, P(D|\alpha I) \]
where $\alpha$ are the parameters one is trying determine, $D$ is the
data and $I$ is the additional information describing the problem.
The quantity $P(\alpha|I)$ is the likelihood, i.e. the probability of
the data given a specific set of parameters, while the left-hand side
is the posterior, i.e. the probability of the parameters given the
data.

\begin{figure}
\center
\epsfig{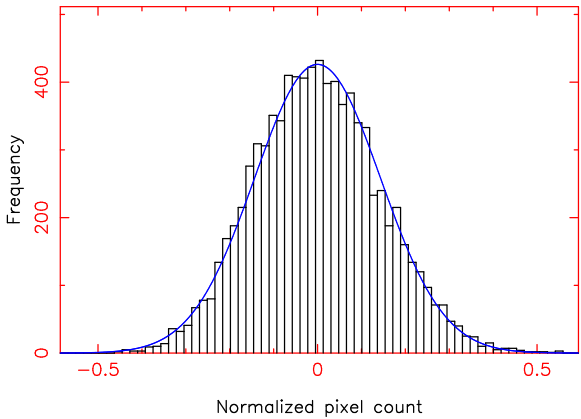}
\caption{Maximum likelihood estimation assumes a Gaussian distribution
for the pixelized data.  This histogram shows the distribution of
values of $x_i$ (see equation \ref{eqdata}) for the NVSS 10 mJy
sample, pixelized using HEALPIX parameter $n_{\rm side} = 32$.  The
overplotted Gaussian curve demonstrates that Gaussianity is a good
approximation for this sample.}
\label{fighist}
\end{figure}

We will assume that the sky is a realization of a stationary Gaussian
process, with an angular power spectrum $C_\ell$.  We assume no
cosmological information about the distribution of the $C_\ell$.  The
rendition of the sky will be a pixelized map, created by binning the
galaxy data in equal-area cells such that the count in the $i$th cell
is $n_i$, effectively constructing a ``temperature map'' of galaxy
surface density.  We performed this task using the HEALPIX software
package (Gorksi, Hivon \& Wandelt 1999; {\tt
http://www.eso.org/science/healpix}).  We chose the HEALPIX
pixelization scheme $n_{\rm side} = 32$, which corresponds to 12,288
pixels over a full sky.  The angular power spectrum may be safely
extracted to multipole $\ell_{\rm max} \approx 2 \times n_{\rm side}$.
We then defined a data vector:
\begin{equation}
x_i = \frac{n_i}{\overline{n}} - 1
\label{eqdata}
\end{equation}
where $\overline{n}$ is the mean count per pixel.  Figure
\ref{fighist} demonstrates that the data vector $x_i$ for the NVSS
sample is well-approximated by a Gaussian distribution, as assumed in
a maximum likelihood analysis.

The covariance matrix $C^T_{ij}$ due to primordial fluctuations is
given by
\[ \langle x_i x_j \rangle \, C^T_{ij} = \sum_\ell \frac{(2\ell+1)}{4\pi} \,
C_\ell \, P_\ell(\cos\theta_{ij}) \]
where $P_\ell$ is the Legendre polynomial and $\theta_{ij}$ is the
angle between pixel pair $(i,j)$.  In order to apply a likelihood
analysis we must also specify a noise covariance matrix $C^N_{ij}$.
We modelled the noise as a Gaussian random process with variance
$1/\overline{n}$, uncorrelated between pixels, such that $C^N_{ij} =
(1/\overline{n}) \, \delta_{ij}$.

The likelihood of the map, with a particular power spectrum $C_\ell$,
is given by
\[ \ln P(C_{\ell}|{\bf x})\propto-\frac{1}{2}({\bf x}^T(C^T+C^N)^{-1}{\bf x})
+Tr[\ln(C^T+C^N)] \] The goal of MLE is to maximize this function, and
the fastest general method is to use Newton-Raphson iteration to find
the zeroes of the derivatives in $\ln P(C_{\ell}|{\bf x})$ with
respect to $C_\ell$.  We used the MADCAP package (Borrill 1999; {\tt
http://www.nersc.gov/$\sim$borrill/cmb/madcap}) to derive the maximum
likelihood banded angular power spectrum from the pixelized galaxy map
and noise matrix.  MADCAP is a parallel implementation of the Bond,
Jaffe \& Knox (1998) maximum-likelihood algorithms for the analysis of
CMB datasets.  We ran the analysis software on the supercomputer
Seaborg, administered by the National Energy Research Scientific
Computing Centre (NERSC) at Lawrence Berkeley National Laboratory,
California.  We again applied equation \ref{eqcldoub} to the MADCAP
results to correct the measured power spectrum for the influence of
multiple-component sources.

Boughn \& Crittenden (2002) also performed a HEALPIX analysis of the
NVSS as part of a cross-correlation analysis with the CMB searching
for evidence of the Integrated Sachs-Wolfe effect.  From the pixelized
map they derived an angular correlation function for the NVSS, which
they used to constrain theoretical models.  We compare their results
with ours in Section \ref{secbias}.

\subsection{Testing the methods}
\label{sectest}

We tested the two methods, direct spherical harmonic and maximum
likelihood estimation, by generating a dipole distribution of $N
\approx 10^5$ sources over the NVSS geometry, where the model dipole
possessed the same amplitude and direction as that detected in the
NVSS (see Blake \& Wall 2002b).  In order to simulate multiple
components, we added companion sources to $\approx 7$ per cent of the
objects, with the same separation distribution as that measured in the
NVSS (Blake \& Wall 2002a).

\begin{figure}
\center
\epsfig{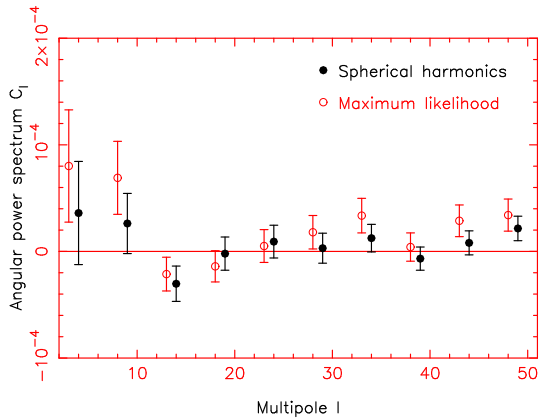}
\caption{The $C_\ell$ spectrum of a model dipole distribution with
added multiple-component sources.  The spectrum is measured using both
spherical harmonic analysis (solid circles) and maximum likelihood
estimation (open circles).  The measurements were averaged into bands
of width $\Delta \ell = 5$, starting from $\ell = 2$.  The results are
consistent with $C_\ell = 0$.}
\label{figcltest}
\end{figure}

The two measurements of the $C_\ell$ spectrum, corrected for multiple
components using equation \ref{eqcldoub}, are plotted in Figure
\ref{figcltest} and are consistent with zero.  The $C_\ell$
measurements have been averaged into bands of width $\Delta \ell = 5$,
starting from $\ell = 2$.  The dipole term $\ell = 1$ is of course
spuriously high, but has a negligible effect on the measured harmonics
at $\ell > 1$ -- the galaxy dipole (unlike the CMB dipole) is only
barely detectable in current surveys (Blake \& Wall 2002b).  The
normalization convention used in Figure \ref{figcltest}, and the
remaining power spectrum plots, is to expand the surface overdensity
$\delta = (\sigma - \sigma_0)/\sigma_0$ in terms of spherical
harmonics.  To convert from the definition of $C_\ell$ in equations
\ref{eqsigylm} and \ref{eqcldef} we simply divide $A_{\ell m}$ by
$\sigma_0$ (in units of sr$^{-1}$) and hence $C_\ell$ by $\sigma_0^2$.

Figure \ref{figcltest} permits a first comparison of the two
independent methods of estimating $C_\ell$.  At low $\ell$, the
variances are in excellent agreement.  As $\ell$ approaches $2 \times
n_{\rm side}$, the variance of the maximum likelihood method begins to
exceed that of the spherical harmonic analysis.  This occurs as the
resolution of the pixelization scheme becomes important: the angular
pixel size is no longer much less than the characteristic angular
scale probed by the $\ell$th multipole.  The variance of the high
$\ell$ bins could be reduced by adopting a finer pixelization, such as
$n_{\rm side} = 64$, with the penalty of a rapidly increasing
requirement of supercomputer time.  Given that the signal in the NVSS
angular power spectrum turns out to be confined to $\ell \la 40$, our
optimum pixelization remains $n_{\rm side} = 32$.  Moreover, higher
pixel resolution would decrease the mean pixel count $\overline{n}$ in
equation \ref{eqdata}, rendering a Gaussian distribution a poorer
approximation for $x_i$.

\section{Results}
\label{secres}

\subsection{The angular power spectrum $C_\ell$}
\label{secmeascl}

\begin{figure}
\center
\epsfig{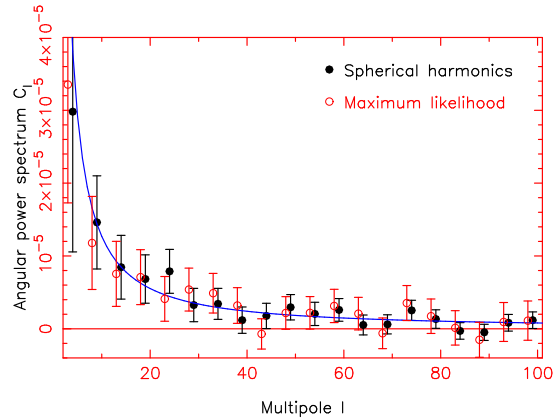}
\caption{The NVSS $C_\ell$ spectrum for a flux-density threshold
$S_{\rm 1.4 \, GHz} = 10$ mJy, determined using two different methods.
The $C_\ell$ measurements are averaged into bins of width $\Delta \ell
= 5$, starting from $\ell = 2$.  The solid circles are the result of a
spherical harmonic analysis and the open circles are derived from a
maximum likelihood analysis.  For clarity, the open circles are offset
by $\Delta \ell = -1$.  The solid line displays the best-fitting
power-law to the spherical harmonic analysis results.}
\label{figcl}
\end{figure}

\begin{figure}
\center
\epsfig{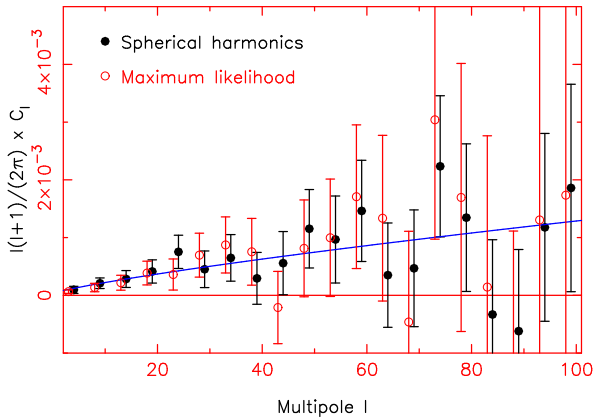}
\caption{The same data points and power-law fit as Figure \ref{figcl},
scaled by a factor $\ell(\ell+1)/2\pi$ to match the usual CMB
convention.}
\label{figclcmb}
\end{figure}

Figure \ref{figcl} plots the NVSS angular power spectrum measured for
flux-density threshold $S_{\rm 1.4 \, GHz} = 10$ mJy using both
spherical harmonic analysis and maximum likelihood estimation.  The
constant offset due to double sources (equation \ref{eqcldoub}) has
been subtracted and the measurements are averaged into bands of width
$\Delta \ell = 5$, starting from $\ell_{\rm min} = 2$.  Measurements
are plotted up to $\ell = 100$, although note that the variance of the
maximum likelihood estimation is increased by pixelization effects
above $\ell \sim 50$, as discussed in Section \ref{sectest}.  Table
\ref{tabcl} lists the plotted data.  Figure \ref{figclcmb} displays
the same data scaled by the usual CMB normalization factor,
$\ell(\ell+1)/2\pi$.

\begin{table}
\center
\caption{Data table of banded NVSS $C_\ell$ values plotted in Figure
\ref{figcl}.  The offset due to double sources is $\Delta C_\ell =
0.24 \times 10^{-5}$.}
\label{tabcl}
\begin{tabular}{cccc}
\hline $\ell_{\rm min}$ & $\ell_{\rm max}$ & Spherical harmonic &
Maximum likelihood \\ & & $C_\ell^{\rm obs} \times 10^5$ &
$C_\ell^{\rm obs} \times 10^5$ \\
\hline
2 & 6 & $2.98 \pm 1.93$ & $3.35 \pm 1.62$ \\
7 & 11 & $1.46 \pm 0.64$ & $1.19 \pm 0.64$ \\
12 & 16 & $0.85 \pm 0.44$ & $0.75 \pm 0.45$ \\
17 & 21 & $0.68 \pm 0.33$ & $0.71 \pm 0.38$ \\
22 & 26 & $0.79 \pm 0.30$ & $0.41 \pm 0.31$ \\
27 & 31 & $0.33 \pm 0.23$ & $0.54 \pm 0.29$ \\
32 & 36 & $0.34 \pm 0.21$ & $0.49 \pm 0.27$ \\
37 & 41 & $0.12 \pm 0.18$ & $0.32 \pm 0.25$ \\
42 & 46 & $0.18 \pm 0.17$ & $-0.07 \pm 0.21$ \\
47 & 51 & $0.30 \pm 0.17$ & $0.22 \pm 0.22$ \\
52 & 56 & $0.21 \pm 0.16$ & $0.22 \pm 0.22$ \\
57 & 61 & $0.26 \pm 0.16$ & $0.31 \pm 0.23$ \\
62 & 66 & $0.05 \pm 0.14$ & $0.21 \pm 0.22$ \\
67 & 71 & $0.06 \pm 0.13$ & $-0.06 \pm 0.21$ \\
72 & 76 & $0.25 \pm 0.14$ & $0.35 \pm 0.24$ \\
77 & 81 & $0.13 \pm 0.13$ & $0.17 \pm 0.24$ \\
82 & 86 & $-0.03 \pm 0.11$ & $0.01 \pm 0.24$ \\
87 & 91 & $-0.05 \pm 0.11$ & $-0.15 \pm 0.24$ \\
92 & 96 & $0.08 \pm 0.11$ & $0.09 \pm 0.27$ \\
97 & 101 & $0.12 \pm 0.11$ & $0.11 \pm 0.27$ \\
\hline
\end{tabular}
\end{table}

We detect clear signal in the NVSS $C_\ell$ spectrum at multipoles
$\ell \la 40$, well-fitted by a power-law $C_\ell = a \, \ell^{-b}$,
where the best-fitting parameters are $a = (2.0 \pm 0.4) \times
10^{-4}$, $b = 1.20 \pm 0.06$ (fitted to the spherical harmonic
analysis result) or $a = (1.6 \pm 0.4) \times 10^{-4}$, $b = 1.12 \pm
0.08$ (fitted to the maximum likelihood method result).  The amplitude
of the angular power spectrum for radio galaxies is two orders of
magnitude smaller than that found for optically-selected galaxies
(e.g. Huterer, Knox \& Nichol 2001); this is readily explained by the
wide redshift range of the NVSS sources, which vastly dilutes the
clustering signal through the superposition of unrelated redshift
slices.  The NVSS signal remains $\sim 5$ orders of magnitude greater
than the CMB $C_\ell$ spectrum over the same multipole range,
reflecting the growth of structure since $z = 1100$.

Given the incomplete sky and finite resolution, the measured $C_\ell$
values are not independent.  However, it was argued in Section
\ref{secestharm} that the correlations between neighbouring power
spectrum measurements are small.  This fact was confirmed by the
maximum likelihood analysis.  The degree of correlation between
neighbouring bins is given by the immediately off-diagonal elements of
the inverse Fisher matrix.  This data is generated by the MADCAP
software, and inspection revealed that the size of the immediately
off-diagonal matrix elements was $\sim 20$ times smaller than that of
the diagonal elements.

\begin{figure}
\center
\epsfig{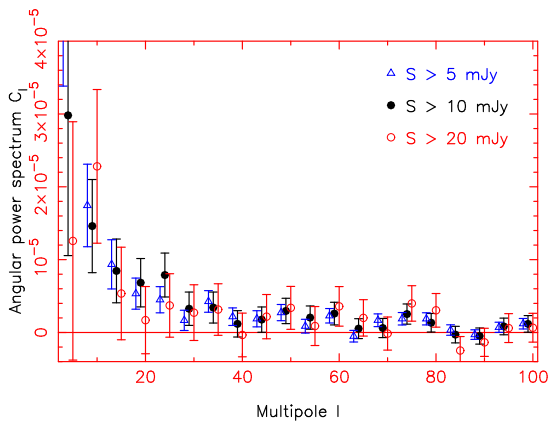}
\caption{The NVSS $C_\ell$ spectrum measured using spherical harmonic
analysis for flux-density thresholds of 5 mJy (triangles, $\sigma_0 =
28.2$ deg$^{-2}$), 10 mJy (solid circles, $\sigma_0 = 16.9$
deg$^{-2}$) and 20 mJy (open circles, $\sigma_0 = 9.5$ deg$^{-2}$).
The 5 mJy points are offset by $\Delta \ell = -1$ and the 20 mJy
points are offset by $\Delta \ell = +1$.}
\label{figclflux}
\end{figure}

Figure \ref{figclflux} compares the angular power spectra measured at
flux-density thresholds 5 mJy, 10 mJy and 20 mJy using spherical
harmonic analysis.  The 5 mJy data may be affected by systematic
surface density gradients.  The results are consistent with an
unchanging underlying power spectrum.  This is not surprising; the
redshift distribution of radio sources does not vary significantly
between 5 mJy and 20 mJy, and the angular correlation function has
been found not to depend on flux density in this range (Blake \& Wall
2002a).

\subsection{Probability distribution of $|A_{\ell m}|^2$}
\label{secalm}

An interesting probe of the galaxy pattern is the distribution of
values of $|A_{\ell m}|^2$ (see Hauser \& Peebles 1973).  These
quantities are measured as part of our spherical harmonic analysis
(Section \ref{secestharm}).  For a random distribution with surface
density $\sigma_0$ over a full sky, the central limit theorem ensures
that the real and imaginary parts of $A_{\ell m} = \sum_i Y_{\ell
m}^*(i)$ are drawn independently from Gaussian distributions such that
the normalization satisfies $|A_{\ell m}|^2 = \sigma_0$.  It is then
easy to show that $x = |A_{\ell m}|^2$ has an exponential probability
distribution for $m \ne 0$:
\begin{equation}
P(x) \, dx = \frac{\exp{(-x/\sigma_0)}}{\sigma_0} \, dx
\label{eqalmdist}
\end{equation}
For a partial sky, $|A_{\ell m}|^2$ is replaced by $|A_{\ell m} -
\sigma_0 I_{\ell m}|^2/J_{\ell m}$ (equation \ref{eqclpeeb}).

\begin{figure}
\center
\epsfig{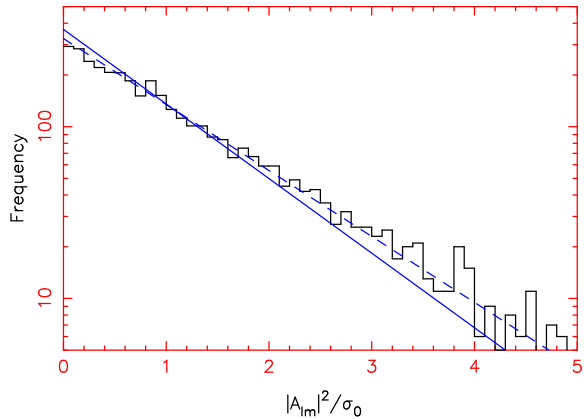}
\caption{The distribution of values of $|A_{\ell m}|^2$ (corrected for
partial sky coverage) for the NVSS for $51 \leq \ell \leq 100$.  The
solid line is the prediction for an unclustered galaxy distribution,
which should be a good approximation given that $C_\ell \approx 0$ for
this range in $\ell$.  However, the slope is a poor fit due to the
influence of multiple-component sources.  When these are taken into
account, the observed distribution is accurately reproduced (the
dashed line).}
\label{figalmdist}
\end{figure}

Figure \ref{figalmdist} plots the distribution of observed values of
$|A_{\ell m} - \sigma_0 I_{\ell m}|^2/J_{\ell m}$.  We restrict this
plot to the multipole range $51 < \ell < 100$: for this range of
$\ell$, Figure \ref{figcl} demonstrates that $C_\ell \approx 0$ and
thus the survey is well-described by a random distribution with
additional multiple components.  For each $\ell$ we included the range
$1 \leq m \leq \ell$ (negative values of $m$ are not independent).
Overplotted on Figure \ref{figalmdist} as the solid line is the
prediction of equation \ref{eqalmdist}.  Multiple components cause the
slope of the observed exponential distribution to be shallower than
this prediction.  Section \ref{secmult} shows that the value of
$<|A_{\ell m} - \sigma_0 I_{\ell m}|^2/J_{\ell m}>$ is increased from
$\sigma_0$ to $(1+2e)\sigma_0$, where $e$ is the fraction of galaxies
split into double sources.  Thus equation \ref{eqalmdist} must be
amended such that $P(x) \propto \exp{[-x/(1+2e)\sigma_0]}$.  Assuming
that $e = 0.07$, this corrected prediction is plotted on Figure
\ref{figalmdist} as the dashed line and provides a very good fit to
the observed distribution.  This is an independent demonstration that
approximately 7 per cent of NVSS galaxies are split into
multiple-component sources.  Figure \ref{figalmdist} also underlines
the fact that the imprint of clustering on the projected radio sky is
very faint.

\subsection{Comparison with $w(\theta)$}

\begin{figure}
\center
\epsfig{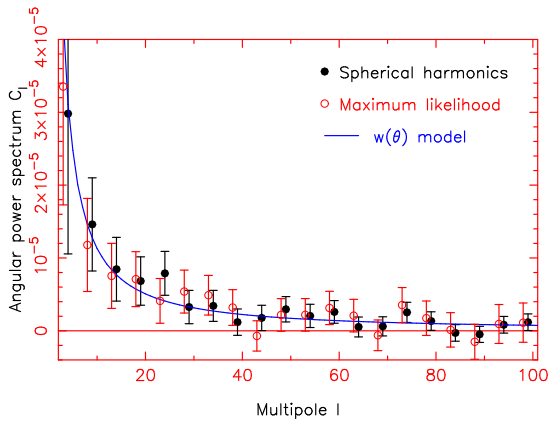}
\caption{The same data points as Figure \ref{figcl}, overplotted with
the prediction of equation \ref{eqwtocl} assuming an angular
correlation function $w(\theta) = (1 \times 10^{-3}) \,
\theta^{-0.8}$.}
\label{figwth}
\end{figure}

The angular correlation function $w(\theta)$ has been measured for the
NVSS by Blake \& Wall (2002a) and Overzier et al. (2003).  It is
well-described by a power-law $w(\theta) \approx (1 \times 10^{-3}) \,
\theta^{-0.8}$ for angles up to a few degrees.  Equation \ref{eqwtocl}
allows us to derive the equivalent $C_\ell$ spectrum if we assume that
this power-law extends to all angular scales.  In Figure \ref{figwth}
we overplot the resulting prediction on the measurements and find an
excellent fit.

This is initially surprising: the angular correlation function is only
measurable at small angles $\theta < 10^\circ$, whereas low multipoles
$\ell$ describe surface fluctuations on rather larger angular scales.
However, in Section \ref{secpk} we establish that the signal in low
multipoles is actually generated at low redshift by spatial
fluctuations on relatively small scales, similar to the spatial scales
which produce the signal in $w(\theta)$.

Of course the comparison is not entirely straight-forward because the
two statistics quantify different properties of the galaxy
distribution.  The value of $C_\ell$ quantifies the amplitude of
fluctuations on the angular scale corresponding to $\ell$.  The value
of $w(\theta)$ is the average of the product of the galaxy overdensity
at any point with the overdensity at a point at angular separation
$\theta$ -- $w(\theta)$ depends on angular fluctuations on all scales.
This is illustrated by the inverse of equation \ref{eqwtocl}:
\[ w(\theta) = \frac{1}{4\pi \, \sigma_0^2} \, \sum_{\ell=1}^\infty (2\ell + 1) \, C_\ell \, P_\ell(\cos{\theta}) \]
For example, a density map constructed using just one multipole $\ell$
possesses a broad angular correlation function.

\section{Relation to the spatial power spectrum}
\label{secpk}

The purpose of this Section is to demonstrate that the observed radio
galaxy $C_\ell$ spectrum can be understood in terms of the current
fiducial cosmological model.  In Section \ref{secbias} we utilize this
framework to determine the linear bias factor of the radio galaxies,
marginalizing over other relevant model parameters.

\subsection{Theory}

The angular power spectrum $C_\ell$ is a projection of the spatial
power spectrum of mass fluctuations at different redshifts, $P(k,z)$,
where $k$ is a co-moving wavenumber.  In linear perturbation theory,
fluctuations with different $k$ evolve independently, scaling with
redshift according to the growth factor $D(z)$.  Under this assumption
we can simply scale the present-day matter power spectrum $P_0(k)$
back with redshift:
\begin{equation}
P(k,z) = P_0(k) \, D(z)^2
\label{eqdz}
\end{equation}
For an $\Omega_m = 1$, $\Omega_\Lambda = 0$ universe, $D(z) =
(1+z)^{-1}$.  For general cosmological parameters we can use the
approximation of Carroll, Press \& Turner (1992).  If linear theory
holds then the angular power spectrum $C_\ell$ can be written in terms
of the present-day matter power spectrum $P_0(k)$ as
\begin{equation}
C_\ell = \int \, P_0(k) \, W_\ell(k) \, dk
\label{eqpktocl}
\end{equation}
(see e.g. Huterer, Knox \& Nichol 2001, Tegmark et al. 2002) where the
kernel $W_\ell(k)$ is given by
\begin{equation}
W_\ell(k) = \frac{2}{\pi} \, \left[ \int_0^\infty j_\ell(u) \, f(u/k)
\, du \right]^2
\label{eqwl}
\end{equation}
Here, $j_\ell$ is a spherical Bessel function and $f(x)$ is a function
which depends on the radial distribution of the sources as
\begin{equation}
f( \, x(z) \, ) = \frac{p(z) \, D(z) \, b(z)}{dx/dz}
\label{eqrad}
\end{equation}
assuming a flat geometry.  $x(z)$ is the co-moving radial co-ordinate
at redshift $z$, $p(z)$ is the redshift probability distribution of
the sources (normalized such that $\int \, p(z) \, dz = 1$) and $b(z)$
is a linear bias factor which relates the clustering of galaxies to
clustering of the underlying mass:
\[ P_{\rm gal}(k,z) = b(z)^2 P_{\rm mass}(k,z) \]
For the purposes of this analysis we assumed that the linear bias does
not evolve with epoch and may be represented by a constant bias factor
$b(z) = b_0$.

A useful approximation for the spherical Bessel function (which gets
better as $\ell$ gets larger) is $j_\ell(x) \approx (\pi/2\ell)^{1/2}
\, \delta(x - \ell)$.  In this approximation,
\begin{equation}
W_\ell(k) \approx \frac{1}{\ell} \, f(\ell/k)^2
\label{eqwlapp}
\end{equation}
Thus as $\ell$ is increased, the kernel just translates along the
$k$-axis.  Combining equations \ref{eqpktocl} and \ref{eqwlapp}
produces the approximation
\begin{equation}
C_\ell \approx \frac{1}{\ell} \int \, P_0(k) \, f(\ell/k)^2 \, dk
\label{eqclkapp}
\end{equation}
or, converting this into an integral over radial co-ordinate,
\begin{equation}
C_\ell \approx \int \, P_0(\ell/x) \, x^{-2} \, f(x)^2 \, dx
\label{eqclxapp}
\end{equation}
In Section \ref{secmeascl} we found that the measured NVSS $C_\ell$
spectrum was well-fit by a power-law.  Inspecting equations
\ref{eqclkapp} and \ref{eqclxapp}, a power law for $C_\ell$ can arise
in two ways:
\begin{itemize}
\item If the function $f(x)$ were a power-law $f(x) \propto x^n$, then
equation \ref{eqclkapp} predicts that $C_\ell \propto \ell^{\, 2n-1}$,
regardless of the form of $P_0(k)$.  In particular, if $f(x)$ were
approximately constant over the relevant scales, then $C_\ell \propto
\ell^{-1}$, which is a good description of the observed NVSS
clustering.
\item If the spatial power spectrum were a power-law $P_0(k) \propto
k^n$, then equation \ref{eqclxapp} predicts that $C_\ell \propto
\ell^n$, regardless of the form of the radial distribution $p(z)$ (see
also Table \ref{tabclus}).  The result $C_\ell \propto \ell^n$ differs
from the case of CMB fluctuations, for which $n = 1$ corresponds to
$C_\ell \propto 1/\ell(\ell+1)$ at low $\ell$ -- the well-known
Sachs-Wolfe effect.  This is because the Sachs-Wolfe effect is
sensitive to fluctuations in gravitational potential, whereas we are
probing fluctuations in mass.
\end{itemize}
In Section \ref{secnz} we determine that the first of these
interpretations is correct.

\begin{table}
\center
\caption{Functional form of common clustering statistics for pure
power-law clustering parameterized by slope $\gamma$.}
\label{tabclus}
\begin{tabular}{lll}
\hline Statistic & Dependence & Validity \\ \hline Spatial correlation
function & $\xi(r) \propto r^{\, -\gamma}$ & All $r$ \\ Angular
correlation function & $w(\theta) \propto \theta^{\, 1-\gamma}$ &
Small $\theta$ \\ Spatial power spectrum & $P(k) \propto k^{\,
\gamma-3}$ & All $k$ \\ Angular power spectrum & $C_\ell \propto
\ell^{\, \gamma-3}$ & High $l$ \\ \hline
\end{tabular}
\end{table}

\subsection{Modelling the present-day power spectrum}
\label{secpkmod}

We assumed that the primordial matter power spectrum is a featureless
power-law, $P_{\rm prim}(k) = A \, k^n$.  On very large scales, the
only alteration to this spectrum in linear theory will be an amplitude
change due to the growth factor.  However, during the epoch of
radiation domination, growth of fluctuations on scales less than the
horizon scale is suppressed by radiation pressure.  This process is
described by the transfer function $T(k)$, such that the present-day
linear matter power spectrum is given by
\begin{equation}
P_0(k) = P_{\rm prim}(k) \, T(k)^2 = A \, k^n \, T(k)^2
\label{eqpkmod}
\end{equation}
Accurate fitting formulae have been developed for the transfer
function $T(k)$ in terms of the cosmological parameters (Eisenstein \&
Hu 1998), which we employed in our analysis (these fitting formulae
assume adiabatic perturbations).  In our fiducial cosmological model,
we fixed the values of the cosmological parameters at $h = H_0/(100
{\rm \; km \, s^{-1} \, Mpc^{-1}}) = 0.73$, $\Omega_m h^2 = 0.134$ and
$\Omega_b/\Omega_m = 0.17$ (Spergel et al. 2003, Table 7, column 3).
We also chose a primordial spectral index $n = 0.97$ (Spergel et
al. 2003), which is close to the predictions of standard inflationary
models.  We consider the effect of variations in these parameter
values in Section \ref{secbias}.  We assumed that the Universe is
flat, with the remaining energy density provided by a cosmological
constant $\Omega_\Lambda = 1 - \Omega_m$.

There are at least two independent ways of estimating the amplitude
$A$ in equation \ref{eqpkmod}.  Firstly we can use constraints on the
number density of massive clusters at low redshift, expressed in terms
of $\sigma_8$, the rms fluctuation of mass in spheres of radius $R = 8
\, h^{-1}$ Mpc:
\begin{equation}
\sigma_R^2 = \frac{9}{2\pi^2 R^2} \int P_0(k) \, [j_1(kR)]^2 \, dk
\label{eqsigsq}
\end{equation}
For example, Viana \& Liddle (1999) determined the most likely value
of $\sigma_8$ using this method to be $\sigma_8 = 0.56 \,
\Omega_m^{-0.47}$.  Alternatively, $A$ can be expressed in terms of
the amplitude of fluctuations at the Hubble radius, $\delta_H$, and
constrained by measurements of CMB anisotropies on large angular
scales:
\begin{equation}
A = 2 \pi^2 \delta_H^2 \left( \frac{c}{H_0} \right)^{3+n}
\label{eqdelh}
\end{equation}
For example, Bunn \& White (1997) give the best-fitting constraint on
$\delta_H$ and $n$ for flat models ($\Omega_m + \Omega_\Lambda = 1$)
based on results of the COBE DMR experiment:
\begin{equation}
\delta_H = 1.94 \times 10^{-5} \, \Omega_m^{-0.785-0.05\ln{\Omega_m}}
e^{-0.95(n-1) - 0.169(n-1)^2}
\label{eqcobe}
\end{equation}
with a maximum $1\sigma$ statistical uncertainty of 7 per cent.  For
our fiducial cosmological parameters, a (reasonably) consistent
cosmology is produced if $\sigma_8 = 1$ (i.e. $\delta_H = 4.7 \times
10^{-5}$).  We assumed this fiducial normalization in our model,
noting that the value of $\sigma_8$ is in fact degenerate with the
amplitude of a constant linear bias factor $b(z) = b_0$.

\begin{figure}
\center
\epsfig{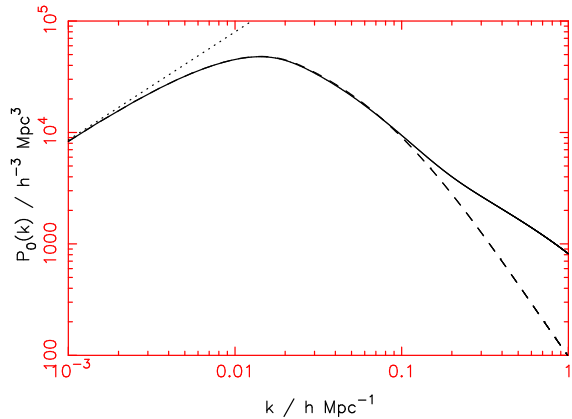}
\caption{The fiducial model present-day matter power spectrum $P_0(k)$
used in this investigation (solid line).  This is modified from a
primordial featureless power-law (dotted line) using the transfer
function of Eisenstein \& Hu (1998), and from the linear power
spectrum (dashed line) using the non-linear prescription of Peacock \&
Dodds (1996).  The assumed cosmological parameters were $h = 0.73$,
$\Omega_m h^2 = 0.134$, $\Omega_b/\Omega_m = 0.17$, $n = 0.97$ and
$\sigma_{8,{\rm lin}} = 1$.}
\label{figpkmod}
\end{figure}

Equation \ref{eqpkmod} is only valid ignoring non-linear effects,
which will boost the value of $P(k)$ on small scales as modes commence
non-linear collapse.  We incorporated non-linear corrections using the
fitting formula provided by Peacock \& Dodds (1996).  The resulting
model power spectrum is shown in Figure \ref{figpkmod}.  Strictly,
this modification violates the assumption of linear evolution implicit
in equation \ref{eqdz}.  However, this is not significant in our
analysis because the small scales for which non-linear evolution is
important are only significant in the projection at $z \approx 0$.

\subsection{Modelling the radial distribution of NVSS sources}
\label{secnz}

The projection of the spatial power spectrum onto the sky depends on
the radial distribution of the sources under consideration, which may
be deduced from their redshift distribution.  We only need to know the
probability distribution $p(z)$ of the sources (equation \ref{eqrad}),
the absolute normalization is not important.  Unfortunately, the
radial distribution of mJy radio sources is not yet accurately known.
The majority of radio galaxies are located at cosmological distances
($z \sim 1$) and their host galaxies are optically very faint.

However, models exist of the radio luminosity function of AGN
(i.e. the co-moving space density of objects as a function of radio
luminosity and redshift), from which the redshift distribution at any
flux-density threshold can be inferred.  Such models have been
published by Dunlop \& Peacock (1990) and Willott et al. (2001).
These luminosity function models are by necessity constrained by
relatively bright radio sources ($S_{\rm 1.4 \, GHz} > 100$ mJy) and
the extrapolation to NVSS flux-density levels must be regarded as very
uncertain.  The Willott model is constrained by a larger number of
spectroscopic redshifts, and the samples of radio sources used provide
fuller coverage of the luminosity-redshift plane (thus the required
extrapolation to $S_{\rm 1.4 \, GHz} = 10$ mJy is less severe).
However, the Willott samples are selected at low frequency (151 and
178 MHz), necessitating a large extrapolation to the NVSS observing
frequency of 1.4 GHz.  The Dunlop \& Peacock models are constrained at
high frequencies, but treat steep-spectrum and flat-spectrum radio
sources as independent populations, which is inconsistent with current
ideas concerning the unification of radio AGN (e.g. Jackson \& Wall
1999).  In addition, they were computed for cosmological parameters
$\Omega_m = 1$, $\Omega_\Lambda = 0$ rather than the currently
favoured ``$\Lambda$CDM'' cosmology.  Furthermore, none of the
aforementioned luminosity function models incorporate starburst
galaxies, which contribute in significant numbers to the radio galaxy
population mix at $z \la 0.1$ for flux-density threshold $S_{\rm 1.4
\, GHz} = 10$ mJy.

Direct measurements of $p(z)$ are currently only achievable at low
redshifts ($z \la 0.2$), where comparison with large optical galaxy
redshift surveys is possible (Sadler et al. 2002; Magliocchetti et
al. 2002).  We matched the NVSS 10 mJy catalogue with the final data
release of the 2dF Galaxy Redshift Survey (2dFGRS, available online at
{\tt http://msowww.anu.edu.au/2dFGRS/}, also see Colless et al. 2001),
in order to estimate $p(z)$ at low redshift.  As our fiducial model,
we fitted the resulting redshift histogram with the simplest possible
function, a constant $p(z) = p_0$ over the range $0 < z < 0.15$.  In
Section \ref{secbias} we include the effect of variations in this
model.  We only considered the redshift range $z < 0.15$ in this
analysis because at $z \approx 0.15$, the (very luminous) optical
counterparts of the 10 mJy NVSS sources begin slipping below the
2dFGRS magnitude threshold, which we verified by plotting the
magnitudes of matched 2dFGRS galaxies against redshift.  Having
determined the value of $p_0$, we created the full redshift
distribution by assigning the remaining probability $1-p_0$ over the
redshift range $z > 0.15$ in proportion to the prediction of the
Dunlop \& Peacock (1990) luminosity function models.  For this
investigation we used the average of the seven models provided by
Dunlop \& Peacock.  Assuming the Willott et al. (2001) radial
distribution for $z > 0.15$ made a negligible difference to the
results (because most of the contribution to the $C_\ell$ spectrum
arises at low redshifts, see Section \ref{secclpred}).

Matching NVSS catalogue entries brighter than $S_{\rm 1.4 \, GHz} =
10$ mJy with the 2dFGRS database yielded $N_{\rm mat} = 546$
identifications with redshifts $z < 0.15$, using matching tolerance 10
arcsec (see Sadler et al. 2002).  We restricted the 2dFGRS sample to
``high quality'' spectra ($Q \ge 3$).  An estimate of the probability
of an NVSS source being located at $z < 0.15$ is
\[ {\rm Prob} = p_0 \, \Delta z = \frac{N_{\rm mat}/A_{\rm
2dF}}{\sigma_{\rm NVSS}} \]
where $\sigma_{\rm NVSS} = 16.9$ deg$^{-2}$ is the surface density of
NVSS sources brighter than 10 mJy, and $A_{\rm 2dF}$ is the 2dFGRS
area under consideration, which (owing to the varying angular
completeness) is not trivial to calculate.  We followed Sadler et
al. (2002) by dividing the number of 2dFGRS galaxies contained in the
NVSS geometry by the 2dFGRS surface density $\sigma_{\rm 2dF} = 180$
deg$^{-2}$, resulting in an effective area $A_{\rm 2dF} = 1214$
deg$^2$.

The result, $p_0 = 0.177$ (per unit redshift), is a significant
underestimate for various reasons:
\begin{itemize}
\item Incompletenesses in the 2dFGRS input catalogue ($5\%$).
\item Input catalogue galaxies unable to be assigned a 2dF
spectrograph fibre ($7\%$).
\item Observed 2dFGRS spectra with insufficient quality ($Q \le 2$) to
determine a redshift ($8\%$).
\item Extended radio sources with catalogue entries located more than
10 arcsec from the optical counterpart ($3\%$).
\end{itemize}
The estimated correction factors in brackets were obtained from
Colless et al. (2001) and from Carole Jackson (priv. comm.).
Multiplying these corrections implies a total incompleteness of
$25\%$, and on this basis we increased the value of $p_0$ to
$0.177/0.75 = 0.237$.

\begin{figure}
\center
\epsfig{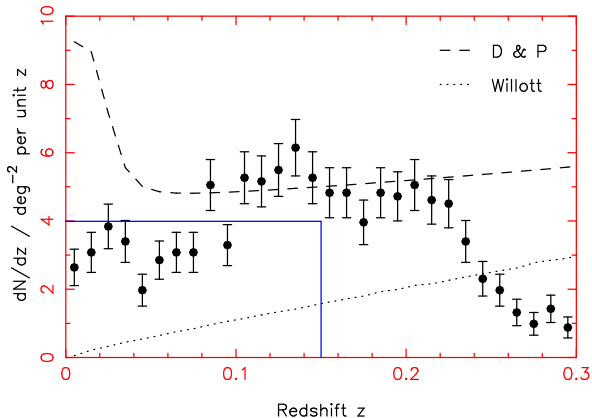}
\caption{The redshift distribution $dN/dz$ of NVSS sources brighter
than 10 mJy.  The data points are measured values resulting from
cross-matching with the 2dF Galaxy Redshift Survey, and the solid line
is the best-fitting constant $dN/dz = p_0 \times \sigma_{\rm NVSS}$
for $0 < z < 0.15$.  We also plot the predictions of two
luminosity-function models: Dunlop \& Peacock (1990, dashed line) and
Willott et al. (2001, dotted line).  The NVSS-2dFGRS matches decline
at $z \ga 0.2$ due to the falling $dN/dz$ of the optical catalogue.}
\label{fignz}
\end{figure}

Figure \ref{fignz} plots 2dFGRS-NVSS matches in redshift bins of width
$\Delta z = 0.01$, together with the low-redshift fit described above.
A constant $p(z)$ is a fairly good approximation at low redshifts ($z
< 0.15$).  This flat distribution arises because the overall redshift
distribution is a sum of that due to AGN and that due to starburst
galaxies; $p(z)$ increases with $z$ for the AGN, but decreases with
$z$ for the starbursters.  Figure \ref{fignz} also displays the
predictions of the luminosity function models of Dunlop \& Peacock
(1990) and Willott (2001).  As explained above, the large
extrapolations involved render these models a poor fit to the redshift
distribution at mJy flux levels, and their use without modification at
low redshift would have caused significant error.

\subsection{Predicting the $C_\ell$ spectrum}
\label{secclpred}

\begin{figure}
\center
\epsfig{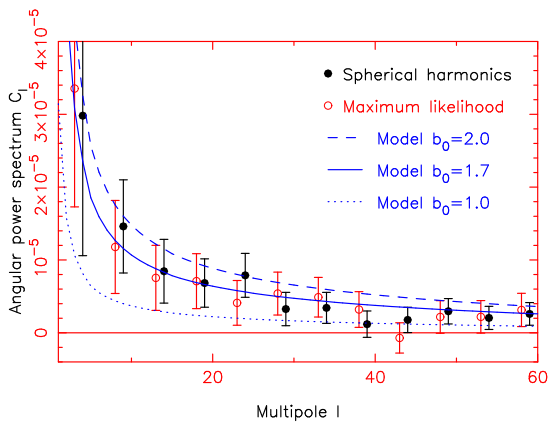}
\caption{The predicted $C_\ell$ spectrum for NVSS sources brighter
than 10 mJy from a reasonable model spatial power spectrum (Section
\ref{secpkmod}) and model redshift distribution (Section \ref{secnz}).
The amplitude of the measurements can only be matched if radio
galaxies are assigned a high bias; model predictions are plotted for
constant linear bias factors $b_0 = 1$ (dotted line), $b_0 = 1.7$
(solid line) and $b_0 = 2$ (dashed line).}
\label{figclmod}
\end{figure}

We used equation \ref{eqpktocl} to predict the $C_\ell$ spectrum from
our fiducial models of the spatial power spectrum (Section
\ref{secpkmod}) and the radial distribution of the sources (Section
\ref{secnz}).  We found that a good match to the measured angular
power spectrum resulted if the NVSS sources were assigned a constant
linear bias factor $b_0 \approx 1.7$ (Figure \ref{figclmod}); $b_0 =
1$ provides a very poor fit to the results.  The bias factor of the
radio galaxies is analyzed more thoroughly in Section \ref{secbias}.

\begin{figure}
\center
\epsfig{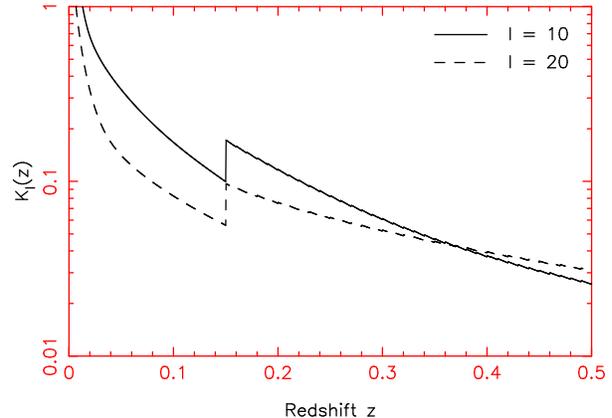}
\caption{Contribution to the model NVSS angular power spectrum as a
function of redshift for multipoles $\ell = 10$ (solid line) and $\ell
= 20$ (dashed line), derived from equation \ref{eqclxapp} using the
assumptions detailed in Sections \ref{secpkmod} and \ref{secnz}.  The
Figure plots $K_\ell(z)$ where $C_\ell = \int K_\ell (z) \, dz$.  The
$y-$axis in the Figure is scaled arbitrarily.  The discontinuity at $z
= 0.15$ results from the break in the model $p(z)$ at this redshift
(see Section \ref{secnz}).}
\label{figkz}
\end{figure}

We note that these measurements of the radio galaxy $C_\ell$ spectrum
at low $\ell$ are {\it not} directly probing the large-scale, small
$k$, region of the power spectrum $P(k)$.  Investigation of the
integrands of equations \ref{eqclkapp} and \ref{eqclxapp} revealed
that the majority of the signal is built up at low redshift, $z \la
0.1$ (see Figure \ref{figkz}), where small-scale spatial power is able
to contribute on large angular scales (i.e. contribute to low
multipoles).  Higher redshift objects principally serve to dilute the
clustering amplitude.  This is unfortunate: the potential of radio
galaxies distributed to $z \sim 1$ to probe {\it directly} the
large-scale power spectrum is forfeited by projection effects.  In
order to realize this potential, we must measure redshifts for the
NVSS sources.  A three-dimensional map extending to $z \sim 1$ would
directly yield $P(k)$ on large scales, defining the ``turn-over''
sketched in Figure \ref{figpkmod}.

\section{Radio galaxy bias factor}
\label{secbias}

The enhanced radio galaxy bias apparent in Figure \ref{figclmod} is
consistent with the nature of AGN host galaxies: optically luminous
ellipticals inhabiting moderate to rich environments.  Similarly high
radio galaxy bias factors have been inferred from measurements of the
spatial power spectrum of low-redshift radio galaxies (Peacock \&
Dodds 1994) and from deprojection of the NVSS angular correlation
function $w(\theta)$ (Blake \& Wall 2002a; Overzier et al. 2003).
Furthermore, Boughn \& Crittenden (2002) cross-correlated NVSS and CMB
overdensities in a search for the Integrated Sachs-Wolfe effect (see
also Nolta et al. 2003).  Their analysis included fitting theoretical
models to the NVSS angular correlation function.  They derived good
fits for a slightly lower linear bias factor than ours, $b_0 = 1.3
\rightarrow 1.6$.  This difference is due to the assumed redshift
distribution.  Boughn \& Crittenden also used the Dunlop \& Peacock
(1990) average model, but we corrected this model using observational
data at low redshifts $z < 0.15$.  As can be seen from Figure
\ref{fignz}, our correction reduces the number of low-redshift
sources, necessitating a higher bias factor to recover the same
angular correlations.

In order to derive a formal confidence interval for the linear bias
parameter $b_0$ we must incorporate the effects of uncertainties in
the underlying model parameters (by marginalizing over those
parameters).  With this in mind, we assumed Gaussian priors for
Hubble's constant $h = 0.73 \, \pm \, 0.03$, for the matter density
$\Omega_m h^2 = 0.134 \, \pm \, 0.006$, and for the primordial
spectral index $n = 0.97 \, \pm \, 0.03$.  The widths of these priors
were inspired by the cosmological parameter analysis combining the
WMAP satellite observations of the CMB and the 2dFGRS galaxy power
spectrum (Spergel et al. 2003, Table 7, column 3) and are a good
representation of our current knowledge of the cosmological model.  In
addition, we considered variations in the model for the radial
distribution of NVSS sources at low redshift (Section \ref{secnz}),
using a more general fitting formula $p(z) = a + b \, z$ to describe
the probability distribution for $z < 0.15$.  For each pair of values
of $(a,b)$ we derived the chi-squared statistic between the model and
the observations, $\chi^2_{p(z)}$, which we converted into an
(unnormalized) probability density $P_{p(z)} \propto
\exp{(-\chi^2_{p(z)}/2)}$.

Our model is thus specified by values of ($b_0$, $h$, $\Omega_m h^2$,
$n$, $a$, $b$) from which we can calculate a model $C_\ell$ spectrum
and hence a chi-squared statistic with the observations,
$\chi^2_{C_\ell}$, corresponding to a probability density $P_{C_\ell}
\propto \exp{(-\chi^2_{C_\ell}/2)}$.  We used the spherical harmonic
estimation of the $C_\ell$ spectrum as the observational data.  After
multiplying $P_{C_\ell}$ by the redshift distribution probability
density $P_{N(z)}(a,b)$ and the Gaussian prior probability densities
for $h$, $\Omega_m h^2$ and $n$, we derived the probability
distribution for $b_0$ by integrating over each of the other
parameters.  We do not marginalize over the normalization of the
matter power spectrum, $\sigma_8$, because this quantity is degenerate
with $b_0$ (using equations \ref{eqrad}, \ref{eqclxapp} and
\ref{eqsigsq}: $C_\ell \propto b_0^2 \, \sigma_8^2$).

\begin{figure}
\center
\epsfig{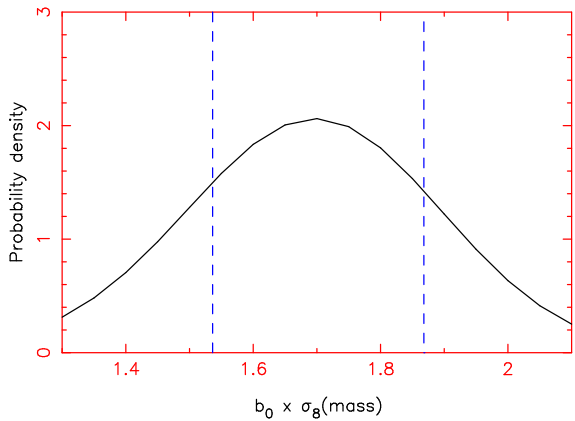}
\caption{Normalized probability distribution for $b_0 \times
\sigma_8$, obtained after marginalizing over the other parameters of
the model as described in Section \ref{secbias}.  The bounds of the
$68\%$ confidence region are marked by the vertical dashed lines.}
\label{figb0}
\end{figure}

The resulting normalized probability distribution for $b_0 \times
\sigma_8$ is displayed in Figure \ref{figb0}, from which we determined
a $68\%$ confidence region $b_0 \, \sigma_8 = 1.53 \rightarrow 1.87$.
When combined with the WMAP determination of $\sigma_8 = 0.9 \pm 0.1$
(Spergel et al. 2003), we infer that $b_0 = 1.89 \pm 0.27$.

\section{Conclusions}

This investigation has measured the angular power spectrum of radio
galaxies for the first time, yielding consistent results through the
application of two independent methods: direct spherical harmonic
analysis and maximum likelihood estimation.  The NVSS covers a
sufficient fraction of sky ($\sim 80$ per cent) that spherical
harmonic analysis is very effective, with minimal correlations amongst
different multipoles.  The form of the $C_\ell$ spectrum can be
reproduced by standard models for the present-day spatial power
spectrum and for the radial distribution of NVSS sources -- provided
that this latter is modified at low redshift through comparison with
optical galaxy redshift surveys.  The results strongly indicate that
radio galaxies possess high bias with respect to matter fluctuations.
A constant linear bias $b_0 \approx 1.7$ permits a good fit, and by
marginalizing over the other parameters of the model we deduce a
$68\%$ confidence interval $b_0 \, \sigma_8 = 1.53 \rightarrow 1.87$
where $\sigma_8$ describes the normalization of the matter power
spectrum.  We find that the majority of the angular power spectrum
signal is generated at low redshifts, $z \la 0.1$.  Therefore, in
order to exploit the potential of radio galaxies to probe spatial
fluctuations on the largest scales, we require individual redshifts
for the NVSS sources.

\section*{Acknowledgments}

We thank Jasper Wall and Steve Rawlings for helpful comments on
earlier drafts of this paper.  We acknowledge valuable discussions
with Carole Jackson concerning cross-matching the NVSS and 2dFGRS
source catalogues.  We are grateful to Sarah Bridle for useful
guidance on marginalizing over the cosmological model.


\begin{thebibliography}{}
\bibitem{1} Baleisis A., Lahav O., Loan A.J., Wall J.V., 1998,
MNRAS, 297, 545
\bibitem{2} Becker R.H., White R.L., Helfand D.J., 1995, ApJ, 450, 559
\bibitem{3} Blake C.A., Wall J.V, 2002a, MNRAS, 329, L37
\bibitem{4} Blake C.A., Wall J.V, 2002b, Nature, 416, 150
\bibitem{5} Bond J.R., Jaffe A.H., Knox L., 1998, PhRvD, 57, 2117
\bibitem{6} Bond J.R., Efstathiou G., Tegmark M., 1997, MNRAS, 291, 33
\bibitem{7} Borrill J., 1999, in {\it Proceedings of the 5th European
SGI/Cray MPP Workshop} ({\tt astro-ph/9911389})
\bibitem{8} Boughn S.P., Crittenden R.G., 2002, PhRvL, 88, 1302
\bibitem{9} Brand K., Rawlings S., Hill G.J., Lacy M., Mitchell E.,
Tufts J., 2003, MNRAS, 344, 283
\bibitem{10} Bunn E.F., White M., 1997, ApJ, 480, 6
\bibitem{11} Carroll S.M., Press W.H., Turner E.L., 1992, ARA\&A, 30,
499
\bibitem{12} Colless M. et al., 2001, MNRAS, 328, 1039
\bibitem{13} Condon J., Cotton W., Greisen E., Yin Q., Perley R.,
Taylor G., Broderick J., 1998, AJ, 115, 1693
\bibitem{14} Cress C., Helfand D., Becker R., Gregg M., White R.,
1996, ApJ, 473, 7
\bibitem{15} Dunlop J.S., Peacock J.A., 1990, MNRAS, 247, 19
\bibitem{16} Efstathiou G., Moody S., 2001, MNRAS, 325, 1603
\bibitem{17} Eisenstein D.J., Hu W., 1998, ApJ, 496, 605
\bibitem{18} Gorski K.M., Hivon E., Wandelt B.D., 1999, in {\it
Proceedings of the MPA/ESO Cosmology Conference ``Evolution of
Large-Scale Structure''}, p.37 ({\tt astro-ph/9812350})
\bibitem{19} Hauser M.G., Peebles P.J.E., 1973, ApJ, 185, 757
\bibitem{20} Hill G.J., Lilly S.J., 1991, ApJ, 367, 1
\bibitem{21} Huterer D., Knox L., Nichol R.C., 2001, ApJ, 555, 547
\bibitem{22} Jackson C.A., Wall J.V., 1999, MNRAS, 304, 160
\bibitem{23} Magliocchetti M., Maddox S., Lahav O., Wall J., 1998,
MNRAS, 300, 257
\bibitem{24} Magliocchetti M. et al., 2002, MNRAS, 333, 100
\bibitem{25} Nolta et al., 2003, ApJ submitted ({\tt
astro-ph/0305097})
\bibitem{26} Overzier R.A., R\"ottgering H.J.A., Rengelink R.B.,
Wilman R.J., 2003, A\&A, 405, 53
\bibitem{27} Peacock J.A., Dodds S.J., 1994, MNRAS, 267, 1020
\bibitem{28} Peacock J.A., Dodds S.J., 1996, MNRAS, 280, 19
\bibitem{29} Peebles P.J.E., 1973, ApJ, 185, 413
\bibitem{30} Sadler E.M. et al., 2002, MNRAS, 329, 227
\bibitem{31} Scott D., Srednicki M., White M., 1994, ApJ, 421, 5
\bibitem{32} Spergel D.N. et al. 2003, ApJS, 148, 175
\bibitem{33} Tegmark M. et al., 2002, ApJ, 571, 191
\bibitem{34} Viana P.T.P., Liddle A.R., 1999, MNRAS, 303, 535
\bibitem{35} Wandelt B.D., Hivon E., Gorski K.M., 2001, PhRvD, 64, 3003
\bibitem{36} Willott C.J., Rawlings S., Blundell K.M., Lacy M., Eales
S.A., 2001, MNRAS, 322, 536
\bibitem{37} Wright E.L., Smoot G.F., Bennett C.L., Lubin P.M., 1994,
ApJ, 436, 441
\end{thebibliography}
\end{document}